\documentclass{article}
\usepackage[a4paper, total={6in, 9.5in}]{geometry}
\usepackage{graphicx}
\usepackage[font=small,labelfont=bf]{caption}
\usepackage[skip=0pt,labelfont=bf]{subcaption}
\usepackage[backend=biber,style=phys,sorting=none]{biblatex}
\usepackage{multicol}
\usepackage[table,dvipsnames]{xcolor}
\usepackage{wrapfig}
\newenvironment{Figure}
  {\par\medskip\noindent\minipage{\linewidth}}
  {\endminipage\par\medskip}
\usepackage{multirow}
\usepackage{array}
\usepackage{hyperref}
\hypersetup{
    colorlinks=true,
    citecolor=OliveGreen,
    linkcolor=BrickRed,
    urlcolor=Blue
    }

\title{\textbf{Benchmarking the human brain against computational architectures}}
\author{Céline van Valkenhoef$^1$\footnote{\href{mailto:celine.vanvalkenhoef@univie.ac.at}{celine.vanvalkenhoef@univie.ac.at}} \and Catherine Schuman$^2$ \and Philip Walther$^{1,3}$\footnote{\href{mailto:philip.walther@univie.ac.at}{philip.walther@univie.ac.at}}
}
\date{\textit{\small
    $^1$University of Vienna, Faculty of Physics and Vienna Center for Quantum Science and Technology (VCQ), Vienna, Austria\\
    $^2$Department of Electrical Engineering and Computer Science, University of Tennessee; Knoxville, USA\\
    $^3$University of Vienna, Research Network for Quantum Aspects of Space Time (TURIS), Vienna, Austria\\}\vspace{-3em}}
\begin{document}
\maketitle
\begin{abstract}
The human brain has inspired novel concepts complementary to classical and quantum computing architectures, such as artificial neural networks and neuromorphic computers, but it is not clear how their performances compare. Here we report a new methodological framework for benchmarking cognitive performance based on solving computational problems with increasing problem size. We determine computational efficiencies in experiments with human participants and benchmark these against complexity classes. We show that a neuromorphic architecture with limited field-of-view size and added noise provides a good approximation to our results. The benchmarking also suggests there is no quantum advantage on the scales of human capability compared to the neuromorphic model. Thus, the framework offers unique insights into the computational efficiency of the brain by considering it a black box.
\end{abstract}
\begin{multicols}{2}
\subsection*{Introduction}
The first mathematical model of computation as defined by Alan Turing in 1936 was inspired by human calculators \cite{turing1936computable}. In subsequent decades, the brain continued to serve as an inspiration for the development of computational models and architectures \cite{rosenblatt1958perceptron, mead2020we,mehonic2022brain,furber2016large} and in turn, these technologies were used to model brain function \cite{yang2020artificial, cichy2019deep}. This recurring cycle has led in particular to the development of artificial neural networks \cite{krogh2008artificial} and neuromorphic hardware \cite{markovic2020physics}, on which machine-learning algorithms complete human-like tasks, from image recognition to natural language processing. The architecture and function of neuromorphic computers are based on neurons and synapses in which processing and memory are collocated. Due to the parallel processing and scalability inherent to neuromorphic computing, large numbers of neurons can operate simultaneously \cite{schuman2022opportunities}. Nevertheless, it remains an open question how well the information processing in the human brain performs, and if this can be described within our current frameworks of artificial intelligence, computability and algorithms \cite{sep-church-turing,brooks2012brain,park2013structural}.
\par In order to compare the computational power of different computer architectures, it is useful to study their resource requirements in terms of time and space for solving a given computational problem, without specific assumptions regarding the underlying architecture or any algorithms. Of particular interest is how the resource requirements change when the size of the computational problem, also referred to as its computational complexity, increases \cite{arora2009computational}. This relation can be expressed as a mathematical relationship, which in combination with the type of computational problem and the type of computer that solves the problem determines its computational class. For example, the complexity class ‘polynomial time’ (\textbf{P}) contains all decision problems that can be solved by classical computers in polynomial time, and the ‘non-deterministic polynomial time’ class (\textbf{NP}) contains all decision problems that can be verified, but not necessarily solved, in polynomial time by classical computers.

\begin{figure*}[b]
    \centering
\begin{subfigure}{0.25\linewidth}
    \centering
    \caption{}
    \includegraphics[width=\textwidth]{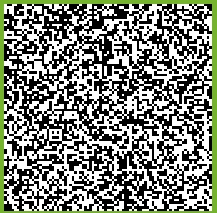}
    \label{fig1a}
\end{subfigure}
\hspace{5em}
\begin{subfigure}{0.25\textwidth}
    \caption{}
    \includegraphics[width=\textwidth]{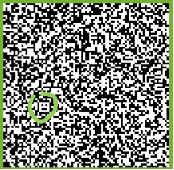}
        \label{fig1b}
\end{subfigure}        
\caption{\textbf{Example of the pattern matching task. (a)}, Example of a pattern-matching task. \textbf{(b)}, As panel a, but with the solution marked.}
\label{fig1}
\end{figure*}

\par Neuromorphic devices process information in distinctly different ways to classical \cite{moore2011nature} and quantum computers \cite{nielsen2002quantum}, which makes it more difficult to define the computational complexity of neuromorphic algorithms \cite{date2021computational}. (A comparison of essential features of classical, quantum and neuromorphic architectures is provided in Table \ref{table 2} in the Supplementary information.) To be more specific, let us consider the resource requirements for solving the computational problem of unstructured search, which aims to find the unique input that matches a particular output value for an unknown function. The simplest algorithm on a classical computer would search all outputs consecutively, thereby taking an amount of time proportional to $n^d$, where $n$ is the length of the output, and $d$ is the number of dimensions of the output. (This scaling is typically denoted O($n^d$).) Quantum computers process information in a fundamentally different manner \cite{einstein1935can,schrodinger1926undulatory}, and algorithms have been developed that exploit these features. In particular, Grover’s algorithm for unstructured search \cite{grover1997quantum} takes O($n^{d/2}$) time to solve the same problem. Neuromorphic devices, finally, process information by exploiting the parallelization of a large number of neurons. In the case of search, the required time steps depend on the number of involved neurons: if the number of neurons is sufficiently large — that is, larger than the problem size — then the neuromorphic device can carry out this task by using only one time step, corresponding to a constant time complexity O(1). This is because all neurons can process an output simultaneously, and only the neuron that matches the presented input fires. If the number of neurons is small, then all outputs have to be searched consecutively and the time complexity will be the same as that of the classical algorithm, O($n^d$). 
\par For an algorithm that works on a given type of computer hardware, we can determine the resource requirements. Vice versa, when we know the resource requirements for a task, we can obtain insights into the efficiency of the algorithms used and the underlying computing architecture. For the brain, we know neither the exact underlying computing architecture nor which, if any, algorithms are used. For example, large-scale brain models are developed to study the structure, function, and disorders of the brain from an interdisciplinary perspective \cite{HBP,brainini}, but these models do not claim to be an exact representation of the brain. On the other hand, basic theories for computational processes in the brain have been proposed \cite{herz2006modeling}, including such in which quantum effects \cite{wigner1995remarks,penrose1990emperor} have a role in neural processing \cite{hameroff2014consciousness,fisher2015quantum,kumar2016possible,adams2020quantum}. 
\par In the present work, we introduce a novel methodology to investigate the human brain: we take a known computational problem, translate it into a task that human participants can solve, and measure the time it takes participants to solve the problem at different problem sizes. We then benchmark the results against different computational complexities, and we formulate a neuromorphic model that approximates these results.
\subsection*{Experimental determination of resource requirements} 
Response time, defined as the time duration between stimulus and response \cite{donders1868over}, can be used to assess the speed of processing in the brain \cite{thorpe1996speed} and is a well-established method in experimental psychology and cognitive science. In these types of experiments, analysis usually focuses on averages, medians, and distributions of the measured response times. In this work, by contrast, we investigate the scaling of response time in the brain of individual study participants with increasing complexity.
\begin{figure*}
\centering
\begin{subfigure}{0.48\textwidth}
    \caption{}
    \includegraphics[width=\textwidth]{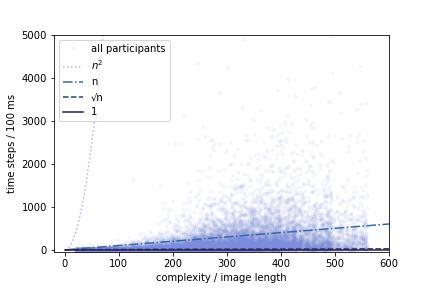}
    \label{fig2a}
\end{subfigure}
\hfill
\begin{subfigure}{0.48\textwidth}
    \caption{}
    \includegraphics[width=\textwidth]{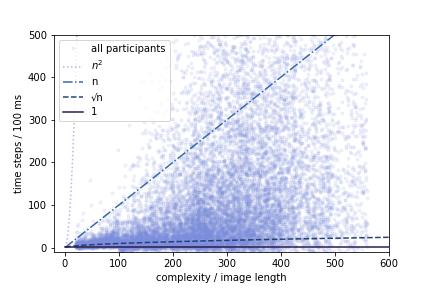}
    \label{fig2b}
\end{subfigure}
\hfill
\begin{subfigure}{0.48\textwidth}
    \caption{}
    \includegraphics[width=\textwidth]{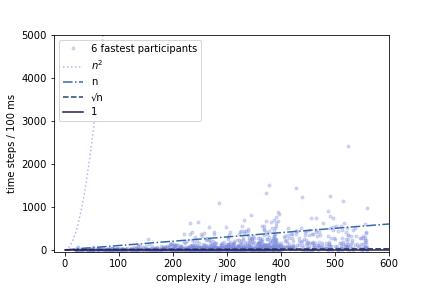}
    \label{fig2c}
\end{subfigure}
\hfill
\begin{subfigure}{0.48\textwidth}
    \caption{}
    \includegraphics[width=\textwidth]{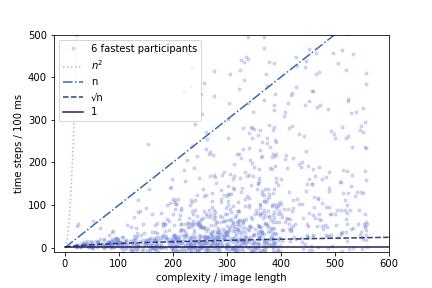}
    \label{fig2d}
\end{subfigure}
\hfill
\begin{subfigure}{0.48\textwidth}
    \caption{}
    \includegraphics[width=\textwidth]{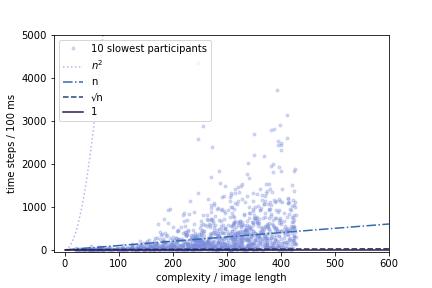}
    \label{fig2e}
\end{subfigure}
\hfill
\begin{subfigure}{0.48\textwidth}
    \caption{}
    \includegraphics[width=\textwidth]{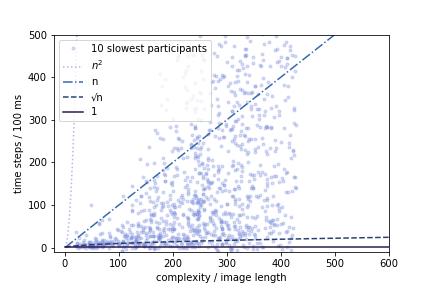}
    \label{fig2f}
\end{subfigure}
\hfill
\begin{subfigure}{0.48\textwidth}
    \caption{}
    \includegraphics[width=\textwidth]{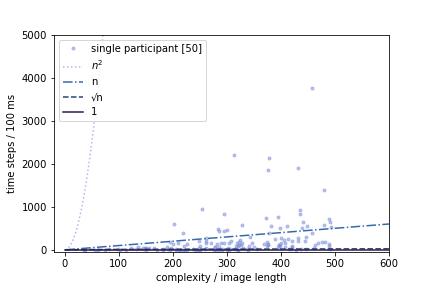}
    \label{fig2g}
\end{subfigure}
\hfill
\begin{subfigure}{0.48\textwidth}
    \caption{}
    \includegraphics[width=\textwidth]{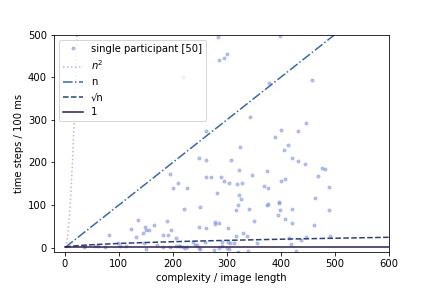}
    \label{fig2h}
\end{subfigure}
\caption{\textbf{Time needed to find a pattern against selected benchmarks. (a, b)}, The results from all participants (a) and an enlarged view of the area of low time consumption (b). \textbf{(c, d)}, Corresponding representations of the results of the six fastest participants. \textbf{(e, f)}, The results of the 10 slowest participants. \textbf{(g, h)}, The results of an average single participant. In all panels, the data are compared with typical scaling relationships.}
\label{fig2}
\end{figure*}
\par The computational problem we explored was that of pattern matching \cite{montanaro2017quantum}, in which the participant had to find a pattern hidden in a scrambled background (see Fig. \ref{fig1} and the Materials and Methods section for details). We posed the problem as a simple, two-dimensional, visual task; as such it does not require any pre-existing skills or knowledge.    
\begin{figure*}[ht]
\centering
\begin{subfigure}{0.49\textwidth}
    \caption{}
    \includegraphics[width=\textwidth]{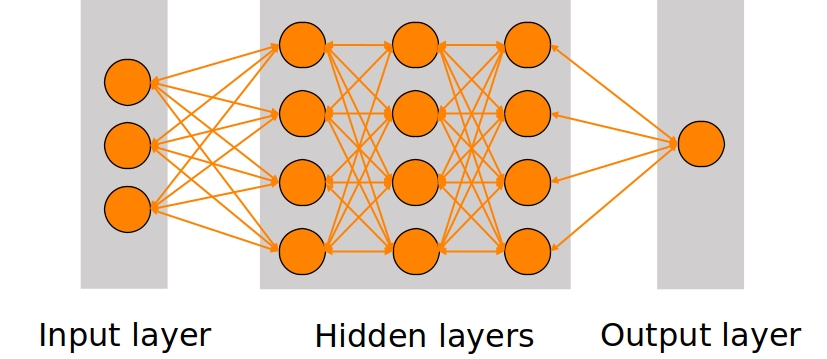}
    \label{fig3a}
\end{subfigure}
\hspace{2em}
\begin{subfigure}{0.29\textwidth}
    \caption{}
    \includegraphics[width=\textwidth]{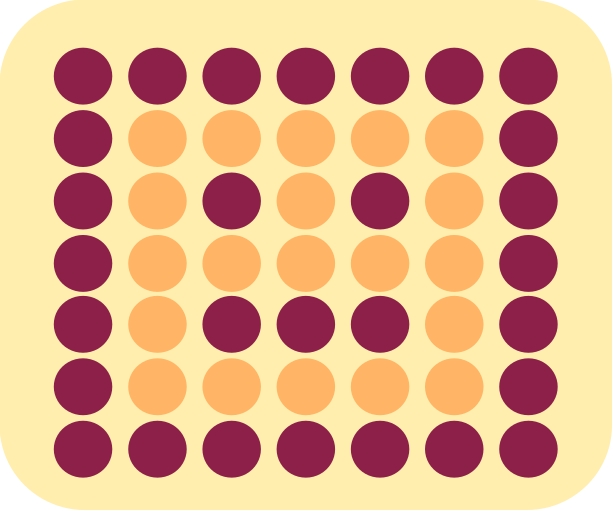}
    \label{fig3b}
\end{subfigure}
\hfill
\begin{subfigure}{0.35\textwidth}
    \caption{}
    \includegraphics[width=\textwidth]{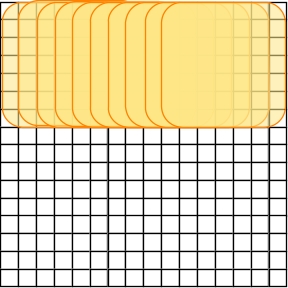}
    \label{fig3c}
\end{subfigure}
\hspace{6em}
\begin{subfigure}{0.35\textwidth}
    \caption{}
    \includegraphics[width=\textwidth]{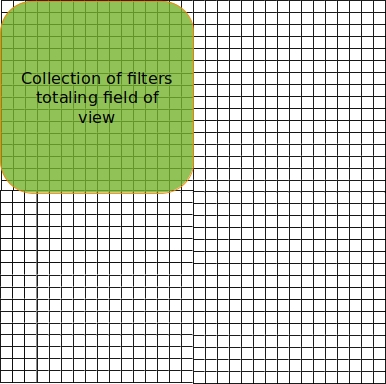}
    \label{fig3d}
\end{subfigure}
       
\caption{\textbf{Neuromorphic model. (a)}, Schematic of the structure of the filter, consisting of an input layer that is fully connected via hidden layers to a single output neuron. \textbf{(b)}, Input layer, whereby the dark and light neurons are programmed to recognize the pattern. \textbf{(c)}, Overlapping filters are positioned over the image. If there are sufficient neurons to allow overlapping filters on the entire image, the pattern matching occurs instantaneously. \textbf{(d)}, A collection of overlapping filters, or ‘field of view’, scans through the image when there are not sufficient neurons.}
\label{fig3}
\end{figure*}
\par Figure \ref{fig2} shows the data collected from the participants, benchmarked against time complexity curves that are characteristic for different efficiencies of various computational architectures. We have chosen to use time steps of 100 milliseconds, as the latency for visual perception in the brain has been found to be on this order of magnitude \cite{thorpe1996speed}. The benchmarked results show that all data points fall below the benchmark $n^2$, a significant proportion of data points fall below the benchmark $n$, and some data points even fall below the benchmark $1$, i.e. constant time. When comparing the results of the slowest and fastest participants, we see that the results of the fastest participants mostly have the benchmark n as upper bound, whereas the slowest participants tend more towards $n^2$. Looking at the results for an average single participant, we see that there is a large variation in processing time for similar image sizes. This is in line with participants’ reported experience that sometimes they could locate the pattern immediately without consciously having to look for it and in other runs, they would have to scan through the image systematically.
\begin{figure*}[ht]
\centering
\begin{subfigure}{0.49\textwidth}
    \caption{}
    \includegraphics[width=\textwidth]{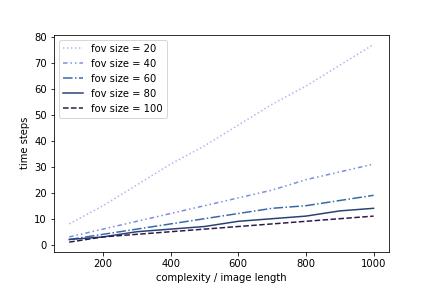}
    \label{fig4a}
\end{subfigure}
\hfill
\begin{subfigure}{0.49\textwidth}
    \caption{}
    \includegraphics[width=\textwidth]{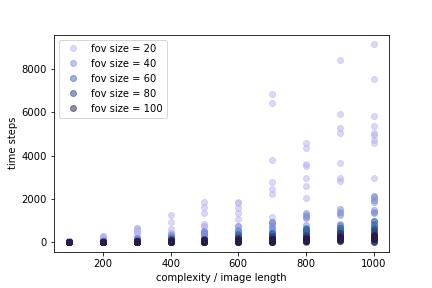}
    \label{fig4b}
\end{subfigure}
\hfill
\begin{subfigure}{0.49\textwidth}
    \caption{}
    \includegraphics[width=\textwidth]{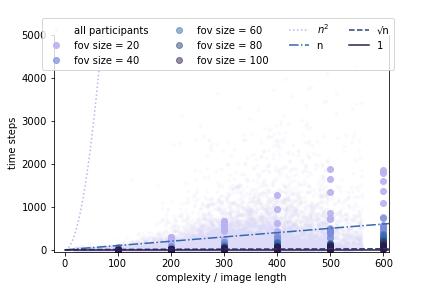}
    \label{fig4c}
\end{subfigure}
\hfill
\begin{subfigure}{0.49\textwidth}
    \caption{}
    \includegraphics[width=\textwidth]{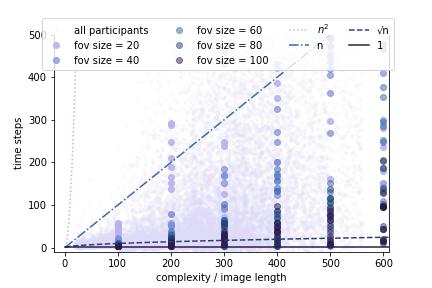}
    \label{fig4d}
\end{subfigure}        
\caption{\textbf{Simulation of time requirements of the neuromorphic search task for different field-of-view (fov) sizes. (a)}, Simulation results without noise. \textbf{(b)}, Simulation results with noise. \textbf{(c)}, Simulation results with noise together with experimental results and benchmarks. \textbf{(d)}, An enlarged view of the low-time-consumption area of panel c. }
\label{fig4}
\end{figure*}
\par The neuromorphic upper bound for two-dimensional search and a small number of neurons, O($n^2$), encloses all data, whereas only a small subset of the fastest-participant data points falls below the bound of a large number of neurons, O($1$). Algorithmically, the neuromorphic model with time complexity O($1$) creates a set of overlapping filters. We consider a neuromorphic computational architecture in which each filter consists of a 7×7 input layer (corresponding to the size of the pattern to be matched) that is fully connected to a single output neuron (see Fig. \ref{fig3a}). The input layer is made up of dark and light neurons with positive and negative weighting respectively, matching the black and white dots of the pattern (Fig. \ref{fig3b}). These dark and light neurons are fully connected to an output neuron that fires when it reaches a threshold value; that is, the weights were set such that the output neurons fire if and only if the exact desired pattern was received as input. Taken together, the overlapping filters cover the entire image simultaneously, leading to instantaneous pattern matching (Fig. \ref{fig3c}). However, the model for O($n^2$) has only sufficient neurons for encoding a single filter, so that scanning through the entire image is needed until the pattern is found. Now consider an intermediate model, in which the number of neurons is limited, but significantly larger than what is needed to encode the pattern. This leads to a ‘field of view’ within which pattern matching is instantaneous, but the field of view has to be scanned through the entire image to find the pattern (Fig. \ref{fig3d}). In this case, a time complexity O($n^2/f^2$) is expected, where $f^2$ is the area of the field of view.
\par Other factors that could also explain some of the observed data and especially the variance observed within the data of a single participant include errors. We found that the location of the pattern within the set — for instance at the borders or more central — did not have a significant effect on the time it took the participants to locate it. 
\par Guided by these findings, we created a neuromorphic model for different field-of-view sizes and applied it to the experimental data (see Fig. \ref{fig4a}). We then added noise to this model to approximate human errors (Fig. \ref{fig4b}–\ref{fig4d}). As expected, for smaller field-of-view sizes, the processing time increases more steeply with increasing problem size, but the shape of the curves does not change. When noise is added, the results show both more variance in time steps per image size, and a large increase in the number of time steps compared to the model without noise. Overall, we find a good qualitative agreement for a model with small field-of-view sizes ($\sim$20) and including noise.
\begin{figure*}[ht]
\centering
\begin{subfigure}{0.49\textwidth}
    \caption{}
    \includegraphics[width=\textwidth]{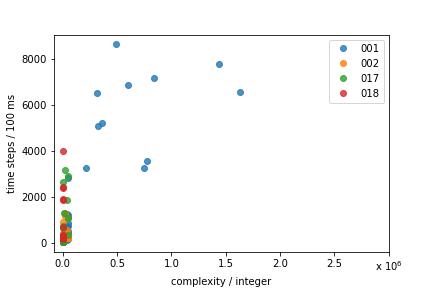}
    \label{fig5a}
\end{subfigure}
\hfill
\begin{subfigure}{0.49\textwidth}
    \caption{}
    \includegraphics[width=\textwidth]{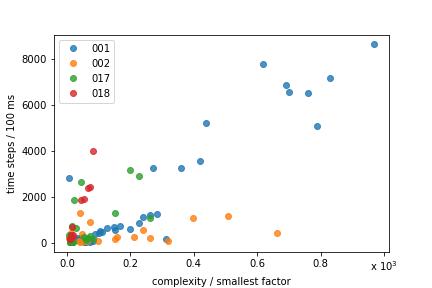}
    \label{fig5b}
\end{subfigure}        
\caption{\textbf{Results of all participants for the factorization task. (a)}, A comparison of processing time against the size of the integer for factorization. \textbf{(b)}, A comparison of processing time against the size of the smallest of the two factors of the solution, where the latter two are taken as a measure of complexity. The highest level of complexity reached was seven-digit numbers. }
\label{fig5}
\end{figure*}
\subsection*{Extension of the method}
So far, we demonstrated a new methodology for investigating the efficiency of the human brain for a specific computational problem, leading to the development of a neuromorphic model that approximates the results. To generalize our method, we both compared our results to other computational models, and we considered an additional computational problem. 
\par To solve the problem of pattern matching using classical and quantum computers, we can use a search algorithm to identify the location of the pattern. Application of a simple classical algorithm in two dimensions leads to a time complexity of O($n^2$), whilst naive application of Grover’s quantum algorithm in two dimensions would have complexity O($n$). Other, faster quantum algorithms for multi-dimensional search may be developed in the future. When we analyse our data in comparison to these benchmarks, we find that the time complexity of the slowest participants is best approximated by O($n^2$), and the time complexity of the fastest participants by O($n$). This provides a first comparison between the performance of the human brain and neuromorphic computing on one hand, and classical and quantum architectures on the other hand. With this, we gain insight into how these different architectures relate in terms of computational complexity.
\par We also extended our method to the problem of prime factorization, whereby the prime numbers are sought whose product equals a given number. Whilst the best known classical algorithms can solve this problem in exponential time, quantum computers can do this, by using Shor’s algorithm, in logarithmic time, O(log $n$), where $n$ is the size of the integer, providing an exponential speed-up \cite{shor1994algorithms}. On the other hand, for humans prime factorization is an extremely difficult task that only few people can solve to a reasonable level of complexity. We recruited the best four expert mental calculators with an interest in prime factorization from the Mental Calculation World Cup \cite{mentalcalc}, as this is a rare skill even among expert mental calculators. These participants were asked to calculate the prime factors of sets of integers with increasing complexity up to seven-digit numbers. Participants reported different methods, including running through a list of memorized prime numbers and checking whether they fit, recognising perfect squares, and trying to find nearby solutions, modular calculation, and ‘just looking at a number and waiting for anything to spring to mind’. 
\par We found a large variation in both processing times and the levels of complexity reached between the individual participants (see Fig. \ref{fig5a}). This is due to both the small sample size and to the specialist skills required to solve the task. There is no clear correlation between processing time and complexity.
\par When taking the smallest of the two factors as a measure of complexity, some correlations seem to appear (see Fig. \ref{fig5b}). While this could suggest that a significant part of the processing time consists of retrieving prime numbers from memory, it is not possible to make any firm conclusions, due to the variation between participants, the small number of participants, and the varying methods they used. We have not attempted to develop a neuromorphic model to simulate these results at this stage. Benchmarking these results against computational algorithms is equally difficult, as the most efficient classical and quantum algorithms work best for much larger numbers. However, these results demonstrate that our method of using reaction-time experiments to collect time-complexity data from human participants is applicable to computational problems other than search tasks. It also suggests that involvement of human memory can be a limiting factor in the applicability of our method.
\subsection*{Discussion}
Over the last centuries, our knowledge of the functioning of the brain has increased enormously. Despite this, due to its immense complexity, there is still a lot to be discovered. Our work makes a significant contribution to the current body of knowledge by demonstrating a new methodology for investigating the efficiency of the human brain for specific computational problems, without specific assumptions regarding the underlying architecture or any algorithms. The application of our ‘black-box’ method to the computational problem of prime factorization highlighted the capabilities and limitations of our method. The insights from applying our new methodology to the search task has led to several important insights. First, the neuromorphic model demonstrates that the performance of the human brain for such a task is similar to that of a classical neural network. Second, benchmarking our results against classical, neuromorphic and quantum algorithms showed that the results of the fastest participants could be approximated by O($n$), representing the quantum search algorithm in two dimensions, whilst also being in the range of the neuromorphic model. This implies that at the ‘human scales’ of our experiment, the neuromorphic model performs as well as the quantum algorithm. Thus, we show that for this search task, the parallel processing of both the human brain and the neuromorphic model is sufficiently powerful to not only outperform the classical algorithm, but also to match the performance of the quantum algorithm. We can explain this finding by noting that time complexities of the quantum algorithm on the one hand and the neuromorphic model and the brain on the other hand scale differently. Due to the powerful processing of the brain and the neuromorphic model, the complexity at which the quantum algorithm becomes more powerful lies outside the ‘human scales’ of our experiment. In other words, for the search task we investigated, there is no practical advantage on ‘human scales’ with respect to time efficiency to use quantum information processing over neuromorphic processing.
\subsection*{Materials and Methods}
\subsubsection*{Search task} 
In our experiments, we used computational complexity to benchmark the efficiency of the human brain against computational architectures in solving specific problems. To obtain processing times for the brain to solve one of the computational problems for a given problem size, we used the method of reaction-time testing. A response time can be seen as a composite measure of multiple processing operations, for example the scanning of a surface, response selection and motor response (see Fig. \ref{fig6}). It is possible to extract processing times from reaction-time tests by posing a ‘baseline test’, in which a simplified form of the task is performed, whereby the part of the task we are interested in — that is, the actual processing time of the problem — is excluded. Then subtracting the time it takes to complete the simplified task from the response time gives the relevant processing time for the task.
\begin{Figure}
\centering
    \includegraphics[width=\textwidth]{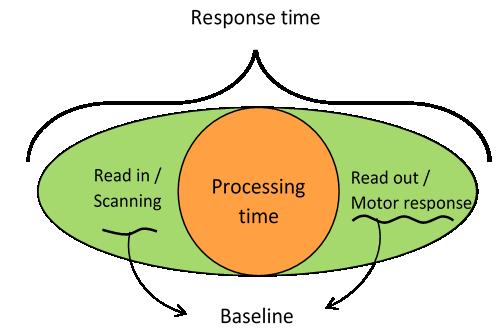}
    \label{fig}
\captionof{figure}{\textbf{Components of response time.} Response time can be considered the sum of processing time and a baseline.}
\label{fig6}
\end{Figure}

\begin{table*}[ht]
\centering
\begin{center}
\newcolumntype{x}{>{\columncolor{gray!60!white!40}} m{2cm}}
\newcolumntype{y}{>{\columncolor{gray!15!white!85}} m{2cm}}
\newcolumntype{z}{>{\columncolor{gray!15!white!85}} m{5cm}}
\begin{tabular}{|x|y|m{.5cm}|x|z|m{1cm}|}
\hline
\rowcolor{gray!60!white!40}
\multicolumn{2}{|l|}{\textbf{Total participants}} & 94 & \multicolumn{2}{|l|} {\textbf{Male/Female}} & 55/39 \\ 
\hline
\textbf{Age range (17-61)}&Under 20&1&\textbf{Education, field}&Anthropology, Art, Culture and Language, and Musicology&13 \\
\hline 
&20-24&37&&Business, Accounting, and Economics&3 \\
\hline
&25-29&34&&Environmental Studies, Sports, Biology, Biomedical and Medical&13\\
\hline
&30-39&17&&Mathematics, Informatics, and Computer Science&8\\
\hline
&40+&5&&None&2\\
\hline
\textbf{Education, highest completed}&None&1&&Physics and Engineering&24\\
\hline
&Secondary&32&&Psychology, Linguistics, and Political Science&29\\
\hline
&Vocational&1&&Teaching&2\\
\hline
&Bachelor&34&&&\\
\hline
&Master&21&&&\\
\hline
&PhD&5&&&\\
\hline  
\end{tabular}
\caption{\textbf{Breakdown of number of participants per characteristic feature.} }
\label{tab1}
\end{center}
\end{table*}
\par We selected the computational problem of pattern matching on the basis that (a) it is a suitable problem to present to participants, (b) memorization does not play a role in solving the problem, and (c) the problem is sufficiently simple so that almost anyone can solve it. This avoids the difficulties associated with the analysis of data where memorization plays a role in problem solving and where there is only a small number of potential participants. The problem of pattern matching can be framed as a simple, two-dimensional, visual task, in which the participant has to find a small pattern hidden in a scrambled background. To aid easy matching and thus avoid the need for memorization, we chose the same pattern of a simple smiley face for all tasks. We specifically chose the pattern to be a smiley face the right way up, as this would be easy for people to identify. As such, we could elicit the best possible response time for a search task to compare with the computational architectures. Both the pattern and the background consisted of black and white dots. The measure of complexity n represents the length of the background, i.e., more complex tasks have a larger background, but there is always just one hidden pattern per task. Participants were told that there was a single pattern hidden in each background, that it was always the same size and always in the same orientation. 
\par Baseline measurements were taken in which the pattern had a different colour than the background, so that participants were able to locate the pattern immediately. Whilst we acknowledge that this is a different task than the main experiment due to the use of colour, we consider this task to be a good approximation for the baseline. This is because the same motor actions need to be performed as in the main task, the same computer programme is used to generate the images and select the pattern, and the identification of the pattern location is almost immediate. Baseline measurements were taken before the main experiment. The length of the background (i.e., the grid length) was taken as the measure of complexity and varied from 20×20 to 565×565 dots. The size of the pattern was 7×7 dots. Most participants completed between 150 and 200 tasks, going from low to high complexity. To avoid fatigue, participants were allowed to take breaks in between tasks when needed. There were 94 participants in total; see Table \ref{tab1} for a breakdown of their characteristic features. There were no correlations between any of these features and the performance of the tasks. 
\subsubsection*{Neuromorphic model}
We developed a model that addresses some of the features that we saw in our data. We created a filter with dark and light neurons (see Fig. \ref{fig3a} and \ref{fig3b}) with positive and negative weighting respectively, that recognizes the pattern when it reaches a threshold value; that is, the weights were set such that the output neurons fire if and only if the exact desired pattern was received as input. When overlapping filters are positioned over the entire image (Fig. \ref{fig3c}), the model can locate the pattern instantaneously (i.e., with complexity O(1)). However, for large image sizes, there might not be a sufficient number of neurons available to cover the entire image. In that case, a model with a limited ‘field of view’ is applied (Fig. \ref{fig3d}). Within this field of view, there are still overlapping filters, thus enabling instantaneous matching, but this field of view is not sufficiently large to cover the entire image. The field of view thus scans through the image, leading to a complexity of O($n^2/f^2$), where f is the length of the field of view and n the length of the image. Finally, we created a second model in which noise was added as an approximation for human errors. In particular, we specified an input noise rate and an output noise rate for our simulation and varied each. The input noise rate (a value between 0 and 1) governed the likelihood that an input neuron that was supposed to fire (because of a black pixel in the image) did fire or a neuron that was not supposed to fire (because of a white pixel in the image) did not fire. Similarly, the output noise rate governed the likelihood that an output neuron that was supposed to fire, did in fact fire. We then applied these two neuromorphic models to the same pattern matching tasks we had given the human participants.
\subsubsection*{Prime factorization task}
In the prime-factorization task, participants were presented with tasks of increasing complexity, meaning that the number they were asked to factorize became increasingly larger. The numbers they had to factorize were presented on a computer screen, where they also had to enter the solutions. Each number that was presented had exactly two prime factors. The response time was measured from the moment the number for factorization appeared on screen to the moment when the participant submitted the solution by pressing the ‘next factorization’ button. Figure \ref{fig7} shows an example of this task.
\begin{Figure}
\centering
\includegraphics[width=\linewidth]{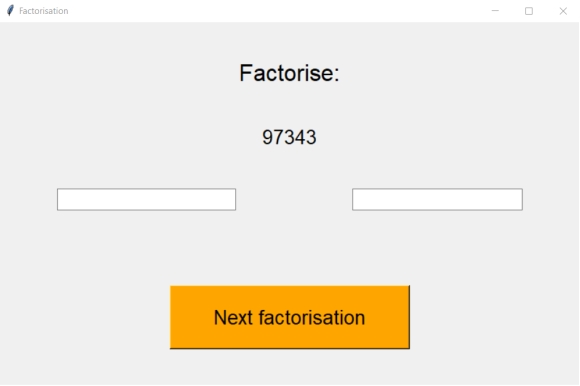}
\captionof{figure}{\textbf{Example of the factorization task.} }
\label{fig7}
\end{Figure}
\par Baseline measurements were taken using small numbers for factorization of maximally three digits, which our expert mental calculators would be expected to solve instantaneously from memory. Participants were not allowed to use note paper or any other tools. The magnitude of the number for factorization was taken as a measure of complexity.
\par There were four participants in total who were expert mental calculators with a special interest in prime factorization. As even among expert mental calculators, the ability to factorize large numbers into primes is not common, and because of the requirement for our research to be able to factorize very large numbers, we were not able to identify any more suitable participants.
\subsubsection*{Acknowledgements}
CvV and PW would like to thank Borivoje Dakic and Claus Lamm for discussions, and the Mental Calculation World Cup organization as well as the University of Vienna’s Cognitive Science Hub for their support in recruiting human research participants. This research received funding from the Austrian Science Fund (FWF) through [F7113] (BeyondC) and the Research Network TURIS. For the purpose of open access, the author has applied a CC BY public copyright licence to any Author Accepted Manuscript version arising from this submission.
\subsubsection*{Author contributions}
CvV performed all experiments and data analysis. CS provided the neuromorphic model and data. All authors contributed to writing the manuscript. PW directed the project.
Ethics declaration
\subsubsection*{Ethics declaration}
The authors declare no competing financial interests.
This project has been approved by the Ethics Committee of the University of Vienna under reference number 00481. Informed consent was obtained from all human research participants.
\printbibliography
\end{multicols}
\newpage
\subsection*{Supplementary information}

\begin{table}[ht]
\centering
\begin{center}
\newcolumntype{s}{>{\columncolor{gray!15!white!85}} m{2.3cm}}
\renewcommand{\arraystretch}{1.5}
\begin{tabular}{|s|m{3.2cm}|m{4.4cm}|m{3.5cm}|}
\rowcolor{gray!60!white!40}
\hline
\textbf{Computer architecture} & \textbf{Neuromorphic} & \textbf{Classical} & \textbf{Quantum}\\
\hline
\textbf{Concept}&
Structure consists of neurons and synapses, in which processing and memory are collocated.& 
Separate CPUs and memory units. Data and programmes are stored in the memory units.& 
Information processing through quantum superposition and quantum entanglement.\\
\hline
\textbf{Encoding}&
Numerical information is encoded in timing, magnitude and shape of spikes.&
Numerical information is encoded in a superposition of states, thus allowing simultaneously 0 and 1.&
Numerical information is binary, i.e. 0 or 1.\\
\hline
\textbf{Processing}&
Due to the parallel processing and scalability, large numbers of neurons can operate simultaneously.&
Calculations are performed on each state separately. The separation of processor and memory causes a slowdown. Calculations can be carried out on N bits simultaneously.&
Due to quantum superposition and quantum entanglement, calculations can be carried out on N quantum bits, and thus on a superposition of 2$^N$ states simultaneously.\\
\hline
\textbf{Applications}&
The structure of neuromorphic computers lends itself to the implementation of machine learning and non-machine learning problems, including graph theory-based algorithms.&
Whilst classical computers can solve any problem that quantum computers can solve, classical computers are generally used to solve problems that fall in complexity class P. Problems in other complexity classes, such as NP, become intractable at increasing complexity.&
The structure of quantum computers lends itself to exploit the structure of some problems, leading to a speed-up in comparison to classical computers in solving these problems.\\
\hline
\end{tabular}
\caption{\textbf{Comparison between classical, quantum and neuromorphic architectures.} }
\label{table 2}
\end{center}
\end{table}
\end{document}